\documentclass{nature}

\usepackage{amsfonts}
\usepackage{amsmath}
\usepackage{amssymb}
\usepackage{revsymb}
\usepackage{graphicx}
\usepackage{mathrsfs}
\usepackage{bm}
\usepackage{color}

\begin{document}

\title{Optomechanically induced transparency of x-rays via optical control}

\author{Wen-Te Liao$^{1,2}$ \& Adriana P\'alffy$^1$}

\maketitle

\begin{affiliations}
\item Max-Planck-Institut f\"ur Kernphysik, Saupfercheckweg 1, D-69117 Heidelberg, Germany
\item Department of Physics, National Central University, 32001 Taoyuan City, Taiwan
\end{affiliations}
\maketitle

\begin{abstract}

The search for new control methods over light-matter interactions is one of the engines that advances fundamental physics and applied science alike. A specific class of light-matter interaction interfaces are setups coupling  photons of distinct frequencies via matter. Such devices, nontrivial in design, could be endowed with multifunctional tasking. 
Here we envisage for the first time an optomechanical system that bridges optical and robust, high-frequency x-ray photons, which are otherwise notoriously difficult to control. 
The x-ray-optical system comprises of an optomechanical cavity and a movable microlever interacting with an optical laser and with x-rays via resonant nuclear scattering. We show that optomechanically induced transparency of a broad range of photons (10 eV-100 keV)  is achievable in this setup, allowing to tune nuclear x-ray absorption spectra via optomechanical control. This paves  ways for metrology applications, e.g., the detection of the $^{229}$Thorium clock transition, and an unprecedentedly precise control of x-rays using optical photons.

\end{abstract}

In cavity optomechanics\cite{Aspelmeyer2014}, the coupling of electromagnetic radiation to mechanical motion degrees of freedom\cite{Brawley2015} can be used to connect quantum system with different resonant frequencies. For instance, via a common movable microlever, an optical cavity can be coupled to a microwave resonator to bridge the two frequency regimes\cite{Barzanjeh2012,Bochmann2013,Andrews2014,Barzanjeh2015}. Going towards shorter photon wavelengths is highly desirable and timely: in addition to improved detection, x-rays are better focusable
and carry much larger momenta, potentially facilitating the entanglement of light and matter at a single-photon level. 
Unfortunately, a direct application of the so-far employed interface concept for a device that  
mediates an optical and an x-ray photon is bound to fail. First, the required high-performance cavities are not available for x-rays. Second, exactly the potentially advantageous high momentum carried by an x-ray photon renders necessary a different paradigm. 
We note here that x-rays  are resonant to transitions in atomic nuclei which can be regarded as x-ray cavities with good quality.  The rapidly developing field of x-ray quantum optics\cite{Suckewer1985,Rocca1994,Lemoff1995,Glover2010,Rohringer2012,Adams2013} has recently reported so far key achievements and promising predictions for the mutual control of x-rays and nuclei\cite{Roehlsberger2010,Roehlsberger2012,Liao2012a,Heeg2013,Liao2014,Liao2014b,Vagizov2014,Liao2015,Heeg2015a,Heeg2015b}.

Here we  present an innovative solution for coupling x-ray quanta to an optomechanical, solid-state device which can serve as a node bridging optical and x-ray photons in a quantum network. We demonstrate theoretically that using resonant interactions of x-rays with nuclear transitions, in conjunction with an optomechanical setup interacting with optical photons, an optical-x-ray interface can be achieved. Such a device would allow to tune x-ray absorption spectra and eventually to shape x-ray wavepackets or spectra for single photons\cite{Perlow1978,Mketchyan1979,Helisto1986,Popov1989,Kocharovskaya1999,Vagizov2006,Vagizov2014} by optomechanical control. The role of the x-ray cavity is here adopted by a nuclear transition with long coherence time that eventually stores the high-frequency photon. Our calculations show that optomechanically induced transparency of x-rays can be achieved in the optical-x-ray interface paving the way for both metrology\cite{Peik2003} and an unprecedently precise control of x-rays using optical photons. In particular, a metrology-relevant application for the nuclear clock transition of $^{229}$Th, which lies in the vacuum ultraviolet (VUV) region, is presented.

\begin{figure}
\vspace{-0.4cm}
  \includegraphics[width=0.5\textwidth]{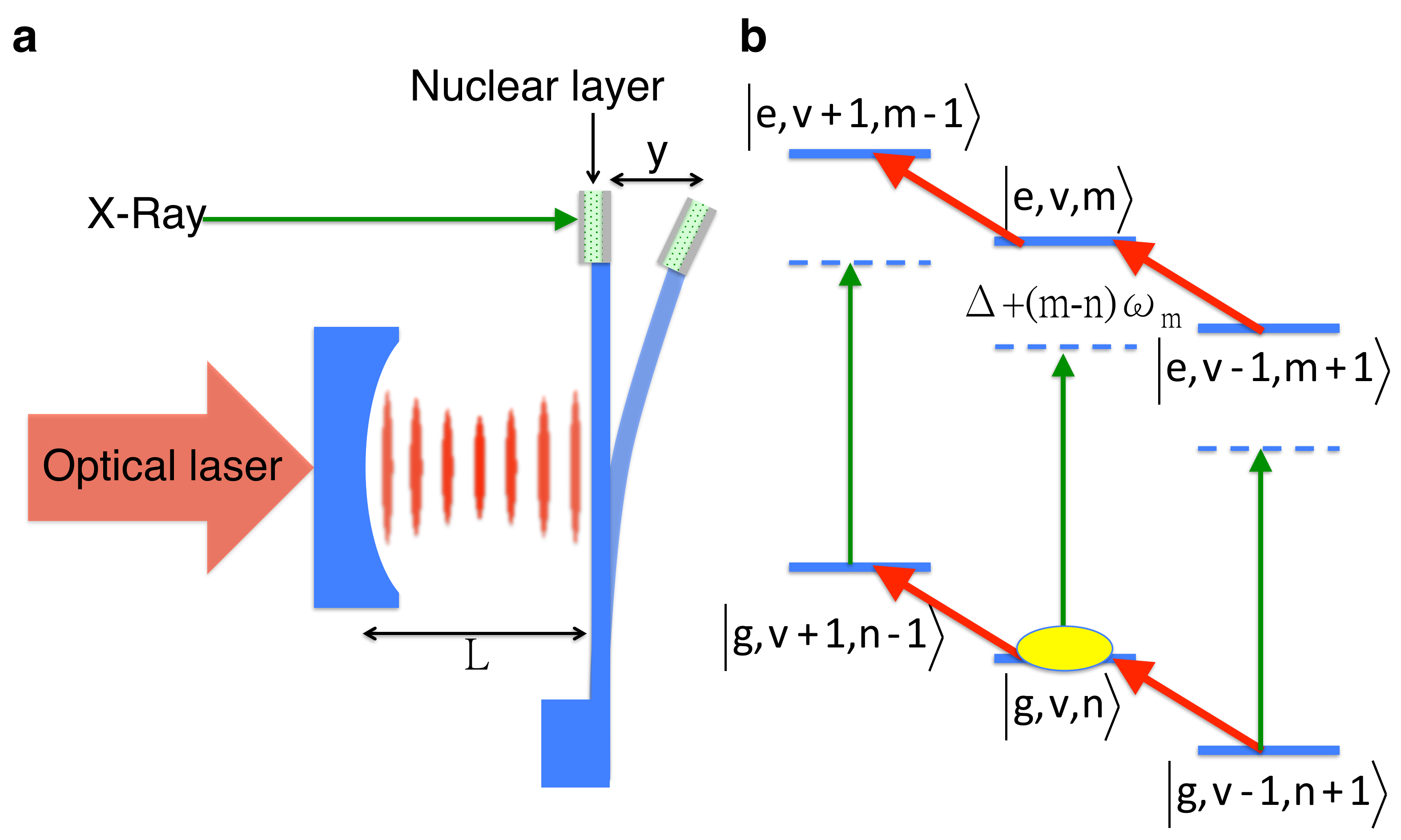}
  \caption{\label{fig1}
Sketch of the optomechanical interface between optical and x-ray photons. {\bf a}, The optical cavity is composed of a fixed mirror and a movable microlever whose oscillating frequency $\omega_m$ can be controlled. A layer containing 
M\"ossbauer nuclei that can resonantly interact with x-rays is embedded in the tip of the microlever. 
{\bf b}, Level scheme of the effective nuclear harmonic oscillator. Lower (upper) three states correspond to the ground (excited) state $g$ ($e$) while $v$ ($n$) denotes the number of fluctuated cavity photons (number of phonons).  Vertical green arrows depict the x-ray absorption by nuclei (with  x-ray detuning  $\Delta$), and red diagonal arrows illustrate the beam splitter interaction between cavity photons and the microlever's mechanical motion. The full yellow ellipse indicates the initial state of the system. }
\end{figure}

The optomechanical-nuclear system under investigation is illustrated in Fig.~\ref{fig1}a. An optomechanical cavity of length $L$ driven by an optical laser has an embedded layer in the tip of the microlever containing M\"ossbauer nuclei that interact with  certain sharply defined x-ray frequencies. 
The nuclei in the layer have a stable or very long-lived ground state, and a first excited state that can be reached by a resonant x-ray M\"ossbauer, i.e., recoilless, transition. 
Typically, this type of nuclear excitation or decay occurs without individual recoil, leading to a coherent scattering in the forward direction\cite{Roehlsberger2004}.  Another type of excitation including the nuclear transition together with the motion of the microlever, i.e.,  phonons, can also be driven by red or blue-detuned x-rays. The nuclear two-level system can be therefore coupled to the mechanical motion of the microlever of mass $M$. 
The term ``phonon'' is used here to describe the vibration of the center of mass of 
the cantilever, visible in the tip displacement $y$.  According to the specifications\cite{Aspelmeyer2014} of various mechanical microlever designs, the phonons in this setup are expected to be in the MHz regime. 
We choose to label the space coordinate with $y$ since the notation $x$ will be used  in the following for the x-ray field.  An effective model of nuclear harmonic oscillator interacting with x-rays can be constructed to describe the hybridization\cite{Shkarin2014,Sete2012b} of the x-ray-nuclei-optomechanical systems. To this end the well-known optomechanical Hamiltonian\cite{Eschner2003,Weis2010,Agarwal2010,Agarwal2012,faust2013,Aspelmeyer2014} is extended to include also the x-ray interaction with the nuclear layer embedded in the tip of the microlever. Since the nuclear transition widths are very narrow ($~10^{-9}-10^{-15}$ eV), we assume that the nuclei interact with a single mode of the x-ray field.

The full Hamiltonian of the system  sketched in Fig.~\ref{fig1}a is a combination of the optomechanical Hamiltonian\cite{Eschner2003,Agarwal2012,Aspelmeyer2014} and nuclear interaction with x-ray photons,
which can be written in the interaction picture and linearized version as (see Methods and Supplementary Information for detailed derivation) as
\begin{eqnarray}
\widehat{H}&=&\hbar\omega_m\widehat{b}^{\dagger}\widehat{b}-\hbar\Delta_c \widehat{a}^{\dagger}\widehat{a}
-\hbar G \left(\widehat{a}^\dagger\widehat{b}+\widehat{a}\widehat{b}^\dagger \right)
\nonumber\\
&+&\hbar\Delta\vert e\rangle\langle e\vert-\frac{\hbar\Omega}{2}\left[ \vert e\rangle\langle g\vert e^{ik_x Y_{\mathrm{ZPF}}\left(\widehat{b}^{\dagger}
+\widehat{b}\right) }+H.c.\right] ,
\label{Hoptomech}
\end{eqnarray}
Here, $\omega_m$ is the optomechanically modified oscillation angular frequency of the microlever,  $\Delta_c$ is the effective optical laser detuning to the cavity frequency obtained after the linearization procedure, 
and $G$ the coupling constant of the system. The operators 
$\widehat{a}^{\dagger}$ ($\widehat{a}$) and  $\widehat{b}^{\dagger}$ ($\widehat{b}$)  act as cavity photon and  phonon creation (annihilation) operators, respectively. As further notations in Eq.~\ref{Hoptomech}, $\Delta=\omega_x-\omega_n$ is the x-ray detuning with $\omega_n$ the nuclear transition angular frequency and $\omega_x$ ($k_x$) is the x-ray angular frequency (wave vector), respectively. $\Omega$ is the Rabi frequency describing the coupling between the nuclear transition currents\cite{Shvydko1996} and the x-ray field, $Y_{\mathrm{ZPF}}$ is the zero-point fluctuation, $\hbar$ the reduced Planck constant, and $e$ and $g$ denote the nuclear excited and ground state, respectively. The linearization procedure leading to the Hamiltonian in Eq.~(\ref{Hoptomech}) was performed in the red-detuned regime, namely, cavity detuning $\Delta_c=-\omega_m$, which results in the so-called ``beam-splitter'' interaction\cite{Aspelmeyer2014} with the optomechanical coupling strength $G$. We use the master equation  
involving the linearized interaction Hamiltonian to determine the dynamics of the interface system and the nuclear x-ray absorption spectra as detailed in Methods.

\begin{figure}
\vspace{-0.4cm}
  \includegraphics[width=0.48\textwidth]{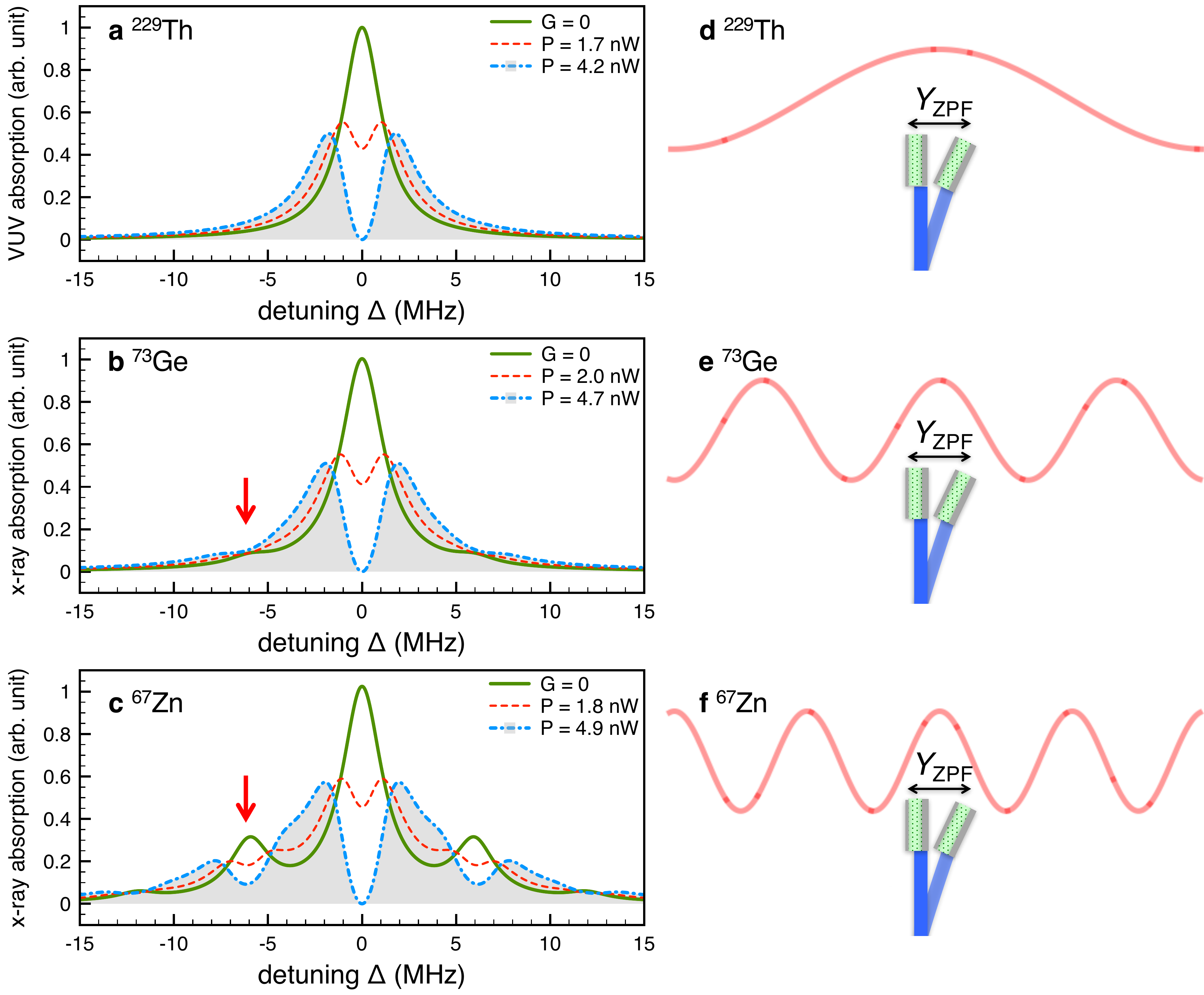}
  \caption{\label{fig2}
 Optomechanically tunable x-ray/VUV absorption spectra and the corresponding ratio of the x-ray wavelength and the zero-point fluctuation $Y_{\mathrm{ZPF}}$.  The microlever has an embedded layer with {\bf a},  $^{229}$Th, {\bf b}, $^{73}$Ge, {\bf c}, $^{67}$Zn nuclei.  Further parameters are taken from  Refs.\cite{Groblacher2009,Aspelmeyer2014} and the phonon number is chosen to be $n=5\times 10^6$. Green solid line illustrates the spectra in the absence of the optomechanical coupling.  Red dashed (blue dashed-dotted) lines show the optomechanically modified spectrum under the action of an optical laser with about $P=$2 nW ($P$=5 nW). Red arrows indicate the first phonon lines. {\bf d-f}, Illustrations of the corresponding ratio of the x-ray wavelength and the zero-point fluctuation $Y_{\mathrm{ZPF}}$ which determine the value of the Lamb-Dicke parameter.
  }
\end{figure}


Figure \ref{fig2} demonstrates the x-ray/VUV absorption spectra  for several nuclear targets, together with an illustration of the corresponding Lamb-Dicke parameter $\eta=k_xY_{\mathrm{ZPF}}$. We consider nuclear transitions from the ground state to the first excited state in $^{229}$Th, $^{73}$Ge and $^{67}$Zn, with the relevant nuclear and optomechanical parameters presented in Table~\ref{table1}. 
The chosen optomechanics setup parameters\cite{Groblacher2009} are $M=0.14$ $\mu$g, the inherent phonon frequency $\omega_0=2\pi\times 0.95$ MHz,  the optomechanical damping rate $\gamma_0=2\pi\times 0.14$ kHz,  the optical cavity decay rate $\kappa=2\pi\times 0.2$ MHz, cavity frequency $\omega_c\sim 10^{15}$ Hz and the optomechanical coupling constant $G_0=\frac{\omega_c}{L}Y_{\mathrm{ZPF}}\sqrt{\overline{n}_\mathrm{cav}}=2\pi\times 3.9$ Hz. These parameters have been experimentally demonstrated\cite{Groblacher2009}. The required optomechanical system is a 25-mm-long Fabry-P\'erot cavity made of a high-reflectivity mirror pad (reflectivity $> 0.99991$) that forms the end-face\cite{Groblacher2009}.
A realistic estimate of the optical thickness values for the nuclear x-ray absorption is presented in the Supplementary Information. 

For a comprehensive explanation, we begin with the case in the absence of optomechanical coupling, i.e., $G=0$.  Green lines in  Fig.~\ref{fig2} illustrate a central nuclear absorption line with detuning $\Delta=0$ corresponding to $m=n$, and sidebands that occur with excitation or decay of phonons in the system $m=n\pm1,n\pm2,\ldots$. The width of the peaks is determined by the value of $s=\Gamma/2+\kappa+\gamma_m$ (see Methods) of the order of MHz, similar in scale with the inhomogeneous broadening 
of the nuclear transition, which we neglect in the following.
In order to resolve the sidebands, 
a constraint has to be imposed on the oscillation frequency of the microlever,  i.e., the microlever frequency $\omega_m>s$, and the Franck-Condon coefficients $\vert F^m_n\vert\gtrsim 0.1$ for at least the first phonon lines $m=n\pm 1$ (see Methods). As a consequence, nuclear species with large Lamb-Dicke parameters $\eta$ allow the observation of x-ray absorption sidebands. For example, compared to $^{229}$Th ($\eta=9.92\times 10^{-9}$) in Fig.~\ref{fig2}a which presents only the zero phonon line, the spectra of $^{73}$Ge ($\eta=1.69\times 10^{-5}$) in Fig.~\ref{fig2}b show an observable sideband as indicated by the red arrow. Moreover,
there are several sidebands appearing in the spectrum of $^{67}$Zn ($\eta=11.87\times 10^{-5}$) depicted in Fig.~\ref{fig2}c.
Further M\"ossbauer nuclei with suitable first excited states whose decay rates are lower than the phonon angular frequency of around 6 MHz are for instance\cite{Roehlsberger2004} $^{45}$Sc, $^{157}$Gd and $^{181}$Ta.

We are now ready to discuss the results including the  optomechanical coupling, $G > 0$, illustrated by the blue and red dashed lines in Fig.~\ref{fig2}. Remarkably, the optomechanical coupling introduces a dip  at the center of each line. As illustrated also in Fig.~\ref{fig1}b, the line splittings are caused by the optomechanical coupling $G$, which links different phonon Fock states via the beam splitter interaction\cite{Aspelmeyer2014}. We stress here that the nuclear x-ray absorption is only modified by the optomechanical coupling and does not have to do with x-ray recoil which is not occuring in our scheme. 
The depth and the spacing of the dips are proportional to the input optical laser power which modifies the strength $G$. Fig.~\ref{fig2} shows that the absorption gradually goes to zero with increasing laser power $P$. The diagonalization of the Hamiltonian shows that the two split peaks around the zero phonon line are approximately positioned at $\Delta=\pm \sqrt{ \left( G\sqrt{m+v+2mv}+s\right)^2-2s^2 }$. These two eigenvalues correspond to transitions between the ground state $\vert g,v,n\rangle$ and the two eigenstates 
$
\sqrt{\frac{\left( 1+m\right)v }{\left( 1+v\right) m}}|e,v-1,m+1\rangle
\mp \sqrt{\frac{m+v+2mv}{\left( 1+v\right) m}}|e,v,m\rangle
+|e,v+1,m-1\rangle
$ (see Methods).
These eigenstates result in an analog of the so-called optomechanically induced transparency\cite{Agarwal2010,Weis2010,Agarwal2012} in the x-ray domain and offer means of controlling x-ray spectra. 
This is a new mechanism compared to typical target vibration experiments of M\"ossbauer samples\cite{Perlow1978,Mketchyan1979,Helisto1986,Popov1989}, in the classical phonon regime.
The width of the splitting indicates that, with sufficient phonon numbers, the compelling optomechanical coupling can be accomplished by an optical laser. 
This feature may render control of x-ray quanta by means of weak optical lasers possible. In order to demonstrate this possibility, laser power parameters of few nW are used in the calculation to implement full transparency of x-rays around the nuclear resonance (see blue dashed-dotted and red dashed lines in Fig.~\ref{fig2}).

Since the natural nuclear linewidths are far more narrow than present x-ray sources, the suitable solution for resolving the phonon sidebands of the keV  x-ray or VUV  resonance energies  is to employ a M\"ossbauer drive setup. $^{67}$Zn M\"ossbauer spectroscopy for instance is a well-established technique  with exceptionally high sensitivity for the gamma-ray energy. This  has been exploited\cite{Zn-Book} for precision measurements of hyperfine interactions
$^{67}$Ga decay schemes, which populate excited states in $^{67}$Zn. The decay cascade will eventually populate the first excited level, which then releases single photons close to the resonance energy of the nuclear layer on the microlever. 
Assuming 50 mCi source activity and a solid angle corresponding to a $20\times20$ $\mu$m$^2$ $^{67}$Zn layer placed  10 cm away, the rate of x-ray photons close to the resonance is approx.~40 Hz. The fine-tuning for matching the exact resonance energy is achieved by means of the Doppler shift using a piezoelectric drive with $\mu$m/s velocities\cite{Zn-Book}.

While the 7.8 eV transition of $^{229}$Th is not traditionally regarded as a M\"ossbauer case, studies have shown that when embedded in VUV-transparent crystals, thorium nuclei are expected to be confined to the Lamb-Dicke regime\cite{Eric2010,Kazakov2012,Liao2012b}. In this regime one expects clear parallels to nuclear forward scattering techniques as known from traditional M\"ossbauer transitions. 
The uniquely low lying state and the very narrow transition width of approx.~$10^{-19}$ eV makes $^{229}$Th a candidate for a stable and accurate nuclear frequency standard\cite{Peik2003}. The most important step in this direction would be a precise measurement of the nuclear transition frequency, at present considered to be 7.8$\pm 0.5$ eV\cite{Beck2007}. However, two major difficulties have been encountered in such measurements. First, the extremely narrow linewidth of $10^{-5}$ Hz makes very difficult both the excitation and the detection of fluorescence for this transition. Second, the isomeric transition has a  disadvantageous signal to background ratio and strong fake signals from the environment have been so far impairing experiments\cite{Stellmer2015,Shaw1999,Utter1999}.

The VUV spectra of $^{229}$Th illustrated in Fig.~\ref{fig2}a reveal that our chip-scale system  could be used to determine the nuclear clock transition energy\cite{Beck2007,Liao2012b,Eric2015}. For this exceptional case with VUV nuclear transition energy, the excitation could be achieved with VUV lasers at present in development\cite{vuv_kddf2}. Two important advantages arise in the VUV-optomechanical interface: (i) the width $s\gg\Gamma$  broadens the VUV absorption linewidth by 10 orders of magnitude, namely, $s\sim 10^{10} \Gamma$, facilitating the excitation and speeding up the nuclear target's decoherence. (ii) The VUV  spectra are optomechanically tunable. This can offer a clear signature of nuclear excitation circumventing false signals which unavoidably appear from either crystal sample\cite{Stellmer2015} or surrounding atmosphere\cite{Shaw1999,Utter1999}.

We have put forward the theoretical formalism for optomechanically induced transparency of x-rays via optical control. In particular, our results show that the induced transparency may be achieved for nuclear transitions, with possible relevance  for metrological studies, e.g., detection of nuclear clock transition. The opposite situation, of x-ray photons controlling the optomechanical setup, may open new possibilities for connecting quantum network devices\cite{Azuma2015} on atomic and mesoscopic  scales.


\subsection{Methods}
 The full Hamiltonian of the system  sketched in Fig.~\ref{fig1}a is a combination of the optomechanical Hamiltonian and nuclear interaction with x-ray photons\cite{Eschner2003,Agarwal2012,Aspelmeyer2014} (see also Supplementary Information),
\begin{eqnarray}
\widehat{H}\label{eq10}
&= &\hbar\omega_0\widehat{b}^\dagger\widehat{b}+\hbar\omega_c\widehat{a}^\dagger\widehat{a} 
-\hbar G_0 \widehat{a}^\dagger \widehat{a} (\widehat{b}^\dagger+\widehat{b})\nonumber\\
&+&\hbar \omega_n |e\rangle\langle e|
+\frac{\hbar\Omega}{2}\left(e^{-i\omega_x t + i k_x Y_{\mathrm{ZPF}}\left(\widehat{b}^{\dagger}+\widehat{b}\right)} \widehat{x}|e\rangle\langle g| \right. \nonumber\\
&+& \left. e^{i\omega_x t - i k_x Y_{\mathrm{ZPF}}\left(\widehat{b}^{\dagger}+\widehat{b}\right)} \widehat{x}^\dagger |g\rangle\langle e|\right)\, .
\end{eqnarray}
Here, $\omega_0$ denotes the inherent phonon, $\omega_c$ the resonant  cavity, and $\omega_n$ the nuclear transition angular frequency, respectively, and 
$\Omega$ is the Rabi frequency describing the coupling between the nuclear transition currents\cite{Shvydko1996} and the x-ray field. The operators 
 $\widehat{x}^{\dagger}$ ($\widehat{x}$) act as x-ray  photon creation (annihilation) operators, respectively.  
The optomechanical coupling constant is given by $G_0=\omega_c Y_{\mathrm{ZPF}}/L$, where $Y_{\mathrm{ZPF}}$ denotes the zero-point fluctuation. 
Typically, the Hamiltonian expression above is transformed in the interaction picture and linearized with respect to the cavity photon number at equilibrium\cite{Agarwal2012,Aspelmeyer2014}, i.e., the balance between external pumping and cavity loss. It is therefore convenient to neglect external cavity driving terms by classical optical fields in the Hamiltonian of the system\cite{Aspelmeyer2014,Agarwal2010,Agarwal2012}. We will see below that one can effectively attribute the modified properties of the system to the new optomechanical coupling constant $G$.
%
By an unitary transformation to the rotating frame\cite{Aspelmeyer2014} (see Supplementary Information), we obtain the Hamiltonian in the interaction picture
\begin{eqnarray}
\widehat{H}&=&\hbar\omega_0\widehat{b}^{\dagger}\widehat{b}-\hbar\Delta_c \widehat{a}^{\dagger}\widehat{a}
-\hbar G_0 \widehat{a}^\dagger \widehat{a} (\widehat{b}^\dagger+\widehat{b})
\nonumber\\
&+&\hbar\Delta\vert e\rangle\langle e\vert-\frac{\hbar\Omega}{2}\left[ \vert e\rangle\langle g\vert e^{ik_x Y_{\mathrm{ZPF}}\left(\widehat{b}^{\dagger}
+\widehat{b}\right) }+H.c.\right]\, ,
\end{eqnarray}
where $\Delta_c=\omega_l-\omega_c$ is the optical laser detuning to the cavity frequency, $\omega_l$ the optical laser angular frequency and $\Delta=\omega_x-\omega_n$ the x-ray detuning. The final step is to linearize the Hamiltonian by performing the transformation $\widehat{a}\rightarrow \sqrt{\overline{n}_{cav}}+\widehat{a}$, where $\overline{n}_{cav}$ is the averaged cavity photon number, and $\widehat{a}$ becomes the photon number fluctuation\cite{Agarwal2012,Aspelmeyer2014}. 
The expression $\overline{n}_{cav}+\langle v\vert \widehat{a}^\dagger \widehat{a}  \vert v\rangle$ gives the photon number of the full cavity field.
We neglect the first order terms of $\widehat{a}^\dagger\widehat{b}^\dagger$ and $\widehat{a}\widehat{b}$ in the rotating wave approximation, and the second order terms proportional to $\widehat{a}^{\dagger}\widehat{a}$. 
The zero order terms $\overline{n}_{\mathrm{cav}} (\widehat{b}^{\dagger}+\widehat{b})$ may be omitted\cite{Aspelmeyer2014} after implementing an averaged cavity length shift $\delta L=\hbar\omega_c \overline{n}_{cav}/(L m \omega_0^2)$ and the
averaged cavity angular frequency shift $\delta \omega_c=\hbar\omega_c^2 \overline{n}_{cav}/(L^2 m \omega_0^2)$, leading to the effective detuning\cite{Aspelmeyer2014}   $\Delta_c\rightarrow \Delta_c+\delta \omega_c$. 
 We focus on the red-detuned regime, namely, cavity detuning $\Delta_c=-\omega_m$, which results in the so-called ``beam-splitter'' interaction\cite{Aspelmeyer2014}. 
  We obtain the linearized Hamiltonian  given in Eq.~(\ref{Hoptomech})
with the new coupling constant $G=G_0\sqrt{\overline{n}_{\mathrm{cav}}}$. The effective phonon angular frequency $\omega_m=\omega_0+\delta\omega_0$ is introduced where $
\delta\omega_0 = 4 G^2\left( \frac{\omega_0}{\kappa^2+16\omega_0^2 }\right)
$ is the optomechanically modified oscillation angular frequency of the microlever\cite{Aspelmeyer2014}. The zero-point fluctuation of the microlever's mechanical motion can then be written as $Y_{\mathrm{ZPF}}=\sqrt{\hbar/\left(2M\omega_m\right)}$.

We use the master equation  $\partial_t\widehat{\rho}=\frac{1}{i\hbar}\left[\widehat{H},\widehat{\rho} \right]+\widehat{\rho}_{dec}$
involving the linearized interaction Hamiltonian to determine the dynamics of the interface system (see Supplementary Information for the explicit form of each matrix). Decoherence processes are described by  $\widehat{\rho}_{dec}$  including the spontaneous nuclear decay characterized by the rate $\Gamma$, the inherent mechanical damping rate of the microlever $\gamma_0$ and  the optical cavity  decay rate $\kappa$.
The density matrix elements $\rho^{\alpha c\mu}_{\beta d\nu}=A^{*}_{\alpha c\mu}A_{\beta d\nu}$ correspond to the state vector 
$|\psi\rangle= 
A_{g v-1 n+1}|g,v-1,n+1\rangle
+A_{g v n}|g,v,n\rangle
+A_{g v+1 n-1}|g,v+1,n-1\rangle
+A_{e v-1 m+1}|e,v-1,m+1\rangle
+A_{e v m}|e,v,m\rangle
+A_{e v+1 m-1}|e,v+1,m-1\rangle$ 
where the system is initially prepared\cite{Groblacher2009,Chen2011} in the nuclear ground state with $\overline{n}_\mathrm{cav}$ fluctuated  cavity photons at the level of $v$ and $n$ phonons $|g,v,n\rangle$, and the nuclear excited state with $m$ phonons $|e,v,m\rangle$ is reached by x-ray absorption, as illustrated in Fig.~\ref{fig1}b.
Four additional states with $n\pm 1$ and $m\pm 1$ phonons are coupled by the beam splitter interaction.
In the red-detuned regime\cite{Aspelmeyer2014} the mechanics of the optically tunable microlever can be described as $\partial_{t}^2y+\gamma_m\partial_{t}y+\omega_m^2 y=0$, where $y(t)$ denotes the displacement of the microlever 
as illustrated in Fig.~\ref{fig1}(a),  and the optomechanical damping rate shift is given by
$
\delta\gamma_0 = 4 G^2\left( \frac{1}{\kappa}-\frac{\kappa}{\kappa^2+16\omega_0^2 }\right)
$. The effective optomechanical damping rate $\gamma_m=\gamma_0+\delta\gamma_0$.
A relevant quantity is the  average number of photons inside the cavity, which depends on the  optical laser power $P$ and is given by\cite{Aspelmeyer2014} $\overline{n}_\mathrm{cav}=\frac{\kappa P}{\hbar\omega_l\left[\left( \omega_l-\omega_c\right)^2+\left( \kappa/2\right)^2\right]}$.

The x-ray absorption spectrum of the interface system is determined by  the off-diagonal terms of the Hamiltonian $\widehat{H}$, i.e., $\langle e,v,m\vert\widehat{H}\vert g,v,n\rangle=\frac{\hbar\Omega}{2}\langle m\vert e^{ik_x Y_{\mathrm{ZPF}}\left(\widehat{b}^{\dagger}+\widehat{b}\right)}\vert n\rangle$. The phase term $\eta=k_x Y_{\mathrm{ZPF}}$ is the so-called Lamb-Dicke parameter, and for $\eta\sqrt{n}<1$, $F^m_n=\langle m\vert e^{i\eta\left(\widehat{b}^{\dagger}+\widehat{b}\right)}\vert n\rangle$ denotes the Franck-Condon coefficient\cite{Eschner2003}
\begin{eqnarray}
F^{m\geq n}_n&=&\frac{\left( i\eta\right)^{\vert m-n\vert} }{\vert m-n\vert !}\sqrt{\frac{m!}{n!}}\, ,
\\ 
F^{m<n}_n&=&\frac{\left( i\eta\right)^{\vert m-n\vert} }{\vert m-n\vert !}\sqrt{\frac{n!}{m!}}\, .
\end{eqnarray}
Typically, only low nuclear excitation is achieved in nuclear scattering with x-rays, such that the master equation in the perturbation region $\Gamma/2+\kappa+\gamma_m > G\gg\Omega$ can be used, corresponding to the stable regime. We note here that  nuclear scattering experiments and simulations have confirmed in  this low excitation regime the validity of the semi-classical limit for x-ray-nucleus interaction\cite{Kong2014}.
The steady state solution reads
\begin{equation}
\rho^{em}_{gn}(\Delta)=
\frac{\Omega\left\lbrace 
F^m_n\left( 2s-\Gamma\right) \left[ is+\Delta-(m-n)\omega_m\right] 
-2iG^2 F^{m-1}_{n-1}\sqrt{mn} 
\right\rbrace  }{2\left( 2 s -\Gamma\right) \left\lbrace  G^2m
+\left[ s-i\left( \Delta-(m-n)\omega_m\right) 
\right]^2\right\rbrace  } 
\label{eq4}
\end{equation}
where the total decoherence rate notation $s=\Gamma/2+\kappa+\gamma_m$ was introduced. By replacing $F^m_n$ with $\vert F^m_n\vert$, the sum of the imaginary part of Eq.~(\ref{eq4}) for corresponding transitions, namely, $\sum_{m=n-6}^{n+6} \mathrm{Im}\left[ \rho^{evm}_{gvn}(\Delta)\right]$, provides the x-ray absorption spectrum. Eq.~(\ref{eq4}) shows that the x-ray absorption is directly dependent on the numbers of photons $\overline{n}_\mathrm{cav}$ and averaged number of phonons $m$ and $n$ and their statistics.

\subsection{Acknowledgements}
The authors would like to thank Markus Aspelmeyer for fruitful discussions. WTL is supported by the Ministry of Science and Technology, Taiwan (Grant No. MOST 105-2112-M-008-001-MY3). WTL is also supported by the National Center for Theoretical Sciences, Taiwan. 
AP gratefully acknowledges funding by the EU FET-Open project 664732. 

\subsection{Author contributions}
WTL and AP contributed equally to this work. WTL performed the numerical calculations. WTL and AP discussed the results and wrote the manuscript text.

\subsection{Competing financial interests}
The authors declare no competing financial interests.

\subsection{Corresponding authors}
Correspondence to wente.liao@g.ncu.edu.tw and palffy@mpi-hd.mpg.de

\subsection{Supplementary information}
See file SupplementaryInformation.pdf.


\begin{table}
\caption{\label{table1} Suitable nuclear M\"ossbauer transitions for the optomechanical control of x-ray absorption.   The nuclear transition energies $E$ and corresponding linewidths $\Gamma$  are given in the first two columns\cite{Roehlsberger2004,Peik2003, Beck2007,Liao2012b}. 
The Lamb-Dicke parameters $\eta$ in the absence of the optical laser obtained for the parameters employed in Fig.~\ref{fig2} are presented in the third column. 
The minimum phonon number $n_{\mathrm{min}}$ for resolving the first phonon line and the valid maximum phonon number $n_{\mathrm{max}}=100 n_{\mathrm{min}}$  are calculated by using $0.1\leq \eta\sqrt{n}< 1$.
}
\center{
\begin{tabular}{lcccc}
\hline
Nucleus     & $E$                 & $\Gamma$              & $\eta$                & $n_{\mathrm{min}}$  \\
            & (keV)               & (MHz)                 & $10^{-5}$             & $10^{7}$            \\ 
\hline \\
$^{45}$Sc   & 12.400              & 2.18$\times 10^{-6}$  &  1.58                 &  4.02                \\
$^{67}$Zn   & 93.312              & 0.08                  & 11.87                 &  0.07 \\ 
$^{73}$Ge   & 13.285              & 0.23                  &  1.69                 &  3.50                \\ 
$^{157}$Gd  & 63.929              & 1.51                  &  8.13                 &  0.15                \\
$^{181}$Ta  & 6.238               & 0.11                  &  0.79                 & 15.88                    \\
$^{229}$Th  & 7.8$\times 10^{-3}$ & $10^{-10}$            &  9.92$\times 10^{-4}$ &  1.02$\times 10^{7}$                                   \\
\hline
\end{tabular}
}
\end{table}



\end{document}